\documentclass[conference]{IEEEtran}
\IEEEoverridecommandlockouts
\usepackage{cite}
\usepackage{amsmath,amssymb,amsfonts}
\usepackage{graphicx}
\usepackage{textcomp}
\usepackage{xcolor}

\usepackage{bm}
\usepackage{amsmath}
\usepackage{amssymb}
\usepackage{amsfonts}
\usepackage{url}
\newcommand{\tfbox}[1]{\fboxsep=0pt\fbox{#1}}
\newcommand{\argmax}{\operatornamewithlimits{argmax}}
\def\vec#1{\mathbf{#1}}

\begin{document}

\title{Finding Appropriate Traffic Regulations\\via Graph Convolutional Networks}

\author{Tomoharu Iwata, Takuma Otsuka, Hitoshi Shimizu, Hiroshi Sawada, Futoshi Naya, Naonori Ueda\\
NTT Communication Science Laboratories}

\maketitle

\begin{abstract}
Appropriate traffic regulations, e.g. planned road closure, are important in congested events. Crowd simulators have been used to find appropriate regulations by simulating multiple scenarios with different regulations. However, this approach requires multiple simulation runs, which are time-consuming. In this paper, we propose a method to learn a function that outputs regulation effects given the current traffic situation as inputs. If the function is learned using the training data of many simulation runs in advance, we can obtain an appropriate regulation efficiently by bypassing simulations for the current situation. We use the graph convolutional networks for modeling the function, which enable us to find regulations even for unseen areas. With the proposed method, we construct a graph for each area, where a node represents a road, and an edge represents the road connection. By running crowd simulations with various regulations on various areas, we generate traffic situations and regulation effects. The graph convolutaional networks are trained to output the regulation effects given the graph with the traffic situation information as inputs. With experiments using real-world road networks and a crowd simulator, we demonstrate that the proposed method can find a road to close that reduces the average time needed to reach the destination.
\end{abstract}


\section{Introduction}

Appropriate traffic regulations for a crowd of people, e.g. planned road closure, 
are important in congested event spaces and urban areas 
as regards improving safety and convenience.
Without regulations, overcrowded traffic might result in fatal accidents caused
by people falling over each other
as well as a reduction in passenger satisfaction due to a prolonged latency time.
Crowd simulation based on multi-agent systems~\cite{braun2003modeling}
%\cite{pelechano2007controlling,narain2009aggregate,braun2003modeling,sung2004scalable} 
has been successfully used
for finding appropriate regulations~\cite{murakami2002multi,almeida2013crowd,DBLP:conf/huc/UedaNSIOS15}.
A crowd simulation under a regulation scenario enables us to calculate
the effect of the regulation, such as the arrival time at the destination and
the degree of congestion.
By calculating the effect of various regulation scenarios, 
we can find an appropriate regulation.
However, this approach requires to conduct many simulations under different regulation scenarios,
which prevents us from finding appropriate regulations to avoid accidents in the current situation.

In this paper, we propose a method to learn a function that 
outputs appropriate regulations
given the current traffic situation as inputs.
If the function is learned using the training data of many simulation runs in advance, 
we can obtain an appropriate regulation by bypassing simulations for the current situation.
For example, we would like to select a road to close to avoid congestion.
Closing a road can reduce congestion by directing the traffic along 
different roads~\cite{youn2008price}. This phenomenon is known as the Braess's paradox.
Also, closing a road in one direction can allow for a better flow in the opposite direction. 
In this case, the input is the number of people on each road, which is represented by a vector,
and the output is the road to be closed.
We can use standard classifiers, such as logistic regression, support vector machines and feed-forward neural networks, to learn the function if we fix the number of roads, or fix the area that we wish to regulate.
However, the learned function cannot be used for other areas
since the input and output size changes depending on the area
and these classifiers require fixed-size inputs and outputs.
Therefore, these standard classifiers are inapplicable to unseen areas.
In addition, learning functions for various areas in advance requires
huge computational time and memory.

To find regulations for various areas with a single function,
we use graph convolutional networks (GCNs)~\cite{schlichtkrull2017modeling}.
The GCN takes variable size feature vectors and graphs as inputs, and outputs values for each node.
With the proposed method, we construct a graph for each area, where a node represents a road, and an edge represents the road connection.
Then, the GCN can employ traffic information for any areas with different numbers of roads in the form of a graph as inputs.
In addition, the GCN can output the road to conduct a regulation for any areas.
By using the GCN with multiple layers, 
we can use the information on roads with the multi-hop distance,
which is transformed so as to be beneficial for finding regulations.

%The remainder of this paper is organalized as follows.
%In Section~\ref{sec:proposed}, we formulate the proposed method.
%In Section~\ref{sec:experiments}, we demonstrate the effectiveness of the proposed method
%by experiments of crosing road regulation.
%In Section~\ref{sec:related}, we briefly review related work.
%Finally, we present concluding remarks and a discussion of future work in Section~\ref{sec:conclusion}.

%Machine learning approaches
%traffic simulator is often used
%for finding good regulation
%traffic simulator is great.
%pedestorian, car, in hall...
%find regulation by what-if similation.
%one direction, stop,....

\section{Proposed method}
\label{sec:proposed}

The proposed method is composed of the following three steps: 
\begin{enumerate}
 \item Generating training data by simulations,
 \item Learning a function using the generated data,
 \item Finding regulations by the learned function.
\end{enumerate}
The proposed method is applicable to pedestrians, cars and bikes
if we use simulators for them.
In the following discussion, we assume for simplicity that 
the regulation procedure is to select a road to close.
However, the proposed method is applicable to other types of regulations
if they can be represented by a graph as discussed
in Section~\ref{sec:extension}. 
 
\subsection{Data generation}

\begin{figure*}[t!]
\centering
{\tabcolsep=0.5em
\begin{tabular}{cccc}
\includegraphics[width=11em]{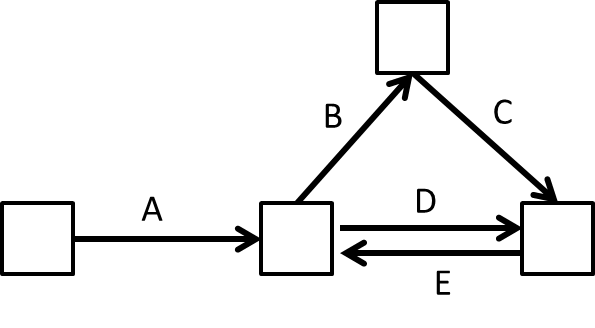}&
\includegraphics[width=10em]{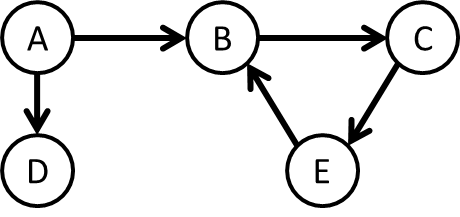}&
\includegraphics[width=11em]{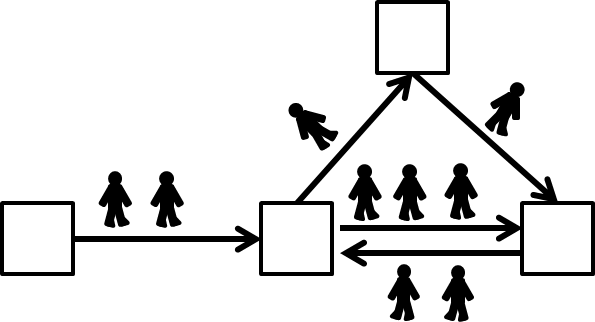}&
\includegraphics[width=11em]{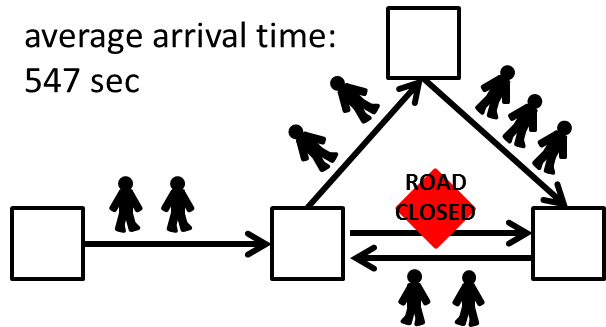} \\
(a) Junction graph&
(b) Road graph&
(c) Calculating traffic situations&
(d) Calculating regulation effect\\
\end{tabular}}
\caption{(a) Junction graph: Each node is a junction or a dead end and the edge is a road.
(b) Road graph: Each node is a road and the edge is the connection between roads.
(c) Calculating traffic situations given the junction graph by simulation.
(d) Calculating the effect when a road `D' is closed in the junction graph by simulation.}
\label{fig:data}
\end{figure*}

First, the proposed method generates
training data to learn a function that outputs 
a way to regulate given road graph and traffic situation.
The training data consist of sets of road graphs, traffic situations and regulation effects.
The data generation procedure is organized in the following four steps.
%\begin{enumerate}
% \item Junction graphs are obtained.
% \item Road graphs are constructed from the junction graph.
% \item Traffic situations are generated by simulations using the road graphs.
% \item Regulation effects are calculated by simulations with various regulation senarios.
%\end{enumerate}

\paragraph{Obtaining junction graphs}

The junction graph is a directed graph,
where each node represents a junction or a dead end, and each edge represents a road.
The junction graphs are used for simulations
as well as for constructing road graphs in the next step.
Figure~\ref{fig:data} (a) shows an example junction graph.

\paragraph{Constructing road graphs}

A road graph is a directed graph,
where each node represents a road,
and each edge represents a connection between roads.
A road graph is constructed from a junction graph.
When the end of a road and the start of another road share a junction,
the corresponding road nodes are connected by a directed edge.
If two roads are the same road running in opposite directions, the nodes are not connected, 
which indicates that roads are not connected by a U-turn.
Figure~\ref{fig:data}(b) shows an example of a road graph constructed from the junction graph
in Figure~\ref{fig:data}(a).
The road graphs are used for the inputs of the function.
Suppose we construct $N$ road graphs, $\{\vec{G}_{n}\}_{n=1}^{N}$, from $N$ junction graphs,
where $\vec{G}_{n}$ is the $n$th road graph with $I_{n}$ nodes.

\paragraph{Calculating traffic situations}
The traffic situation represents the circumstances of each road, such as
the population and the average moving speed.
We generate a traffic situation for each road graph using a crowd simulator.
The crowd simulator calculates the number of pedestrian 
for each road over time
given a junction graph, origin, destination and start time for each pedestrian.
These inputs for the simulator can be set randomly.
For example, when we use the population on each road as the traffic situation, 
it is represented by an $I_{n}$-dimensional vector, $\vec{x}_{n}=\{x_{ni}\}_{i=1}^{I_{n}}$,
where $x_{ni}\in\mathbb{R}$ is the number of pedestrians on road $i$
in graph $n$.
Figure~\ref{fig:data}(c) shows the simulation on the junction graph.

\paragraph{Calculating regulation effects}

The regulation effects for each traffic situation are calculated 
with a simulator.
We save the simulation result when we obtain the traffic situation in the previous step.
Next, we conduct a regulation operation, 
i.e. closing a road in the road graph, 
and run simulation from the saved point.
Then, we calculate the regulation effect, such as 
the arrival time at the destination,
the degree of congestion,
the population that arrived at their destination,
and the number of congested roads.
Figure~\ref{fig:data}(d) shows the simulation result when road `D' is closed.
The regulation effects on graph $n$ 
is represented by a vector $\vec{y}_{n}=\{y_{ni}\}_{i=1}^{I_{n}}$,
where $y_{ni}\in\mathbb{R}$ is the regulation effect value when road $i$ is closed.
We assume that a higher $y_{ni}$ indicates better regulation.

\subsection{Function learning}

Given sets of triplets of road graph, traffic situation and regulation effect,
${\cal D}=\{(\vec{G}_{n},\vec{x}_{n},\vec{y}_{n})\}_{n=1}^{N}$,
we learn a function that outputs the regulation effects
given road graph $\vec{G}$ and traffic situation $\vec{x}$.
We use the graph convolutional networks (GCNs) for the function.

\begin{figure}
\centering
\includegraphics[width=18em]{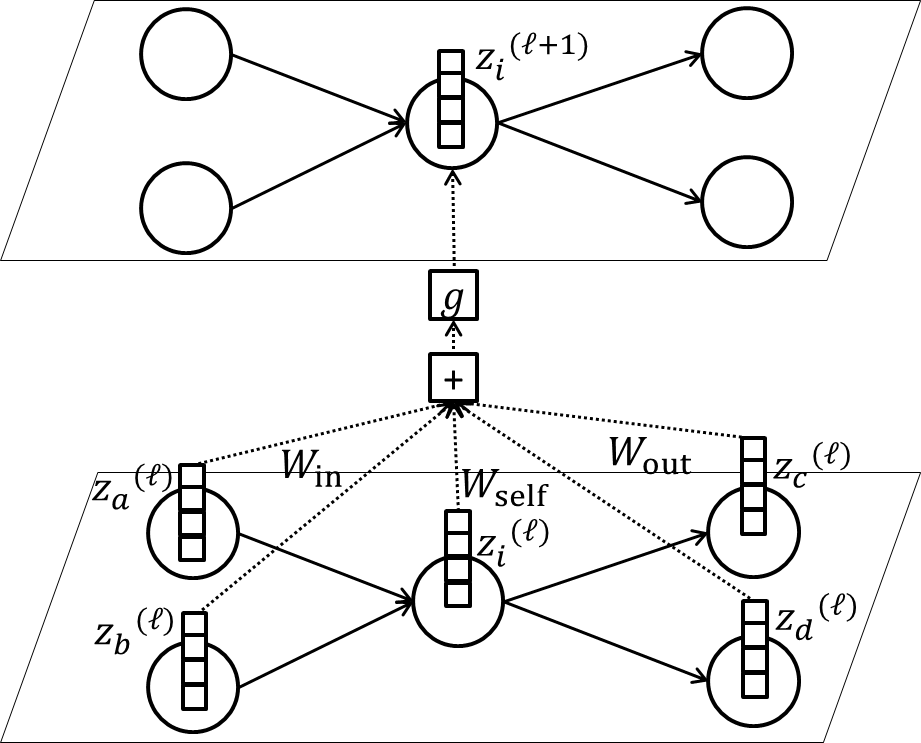}
\caption{Calculation of the hidden states of the next layer, 
$\vec{z}_{i}^{(\ell+1)}$, with the graph convolutional network. 
It is calculated as follows,
1) the hidden states of the incoming nodes, $\vec{z}_{a}^{(\ell)}$, $\vec{z}_{b}^{(\ell)}$, 
the node itself, $\vec{z}_{i}^{(\ell)}$, and the outcoming nodes, $\vec{z}_{c}^{(\ell)}$, $\vec{z}_{d}^{(\ell)}$,
are linearly transformed by $\vec{W}_{\rm in}$, $\vec{W}_{\rm self}$ and $\vec{W}_{\rm out}$, respectively,
2) summed up by `+', and 
3) nonlinearly transformed by activation function $g$.}
\label{fig:gcn}
\end{figure}

With the GCN, each node has $L$-layer hidden states.
Let $\vec{z}_{ni}^{(\ell)}$ be the $\ell$th-layer hidden state of 
the node $i$ of graph $n$.
The $0$th layer corresponds to the input traffic situation itself, $\vec{z}_{ni}^{(0)}=x_{ni}$.
The hidden state of the next layer is calculated 
using the hidden states of its connected nodes and the node itself at the current layer
as follows,
\begin{align}
\vec{z}_{ni}^{(\ell+1)}&=g \Biggl(
\sum_{j\in{\cal I}_{ni}}\frac{1}{|{\cal I}_{ni}|}\vec{W}_{\rm in}^{(\ell)}\vec{z}_{nj}^{(\ell)}
\nonumber\\
&+
\sum_{j\in{\cal O}_{ni}}\frac{1}{|{\cal O}_{ni}|}\vec{W}_{\rm out}^{(\ell)}\vec{z}_{nj}^{(\ell)}+
\vec{W}_{\rm self}^{(\ell)}\vec{z}_{ni}^{(\ell)} \Biggr),
\label{eq:gcn}
\end{align}
where
${\cal I}_{ni}$ is a set of nodes connected to node $i$,
${\cal O}_{ni}$ is a set of nodes connected from node $i$,
which are defined by the road graph $\vec{G}_{n}$,
$\vec{W}_{\rm in}^{(\ell)}\in\mathbb{R}^{H_{\ell+1}\times H_{\ell}}$,
$\vec{W}_{\rm out}^{(\ell)}\in\mathbb{R}^{H_{\ell+1}\times H_{\ell}}$ and
$\vec{W}_{\rm self}^{(\ell)}\in\mathbb{R}^{H_{\ell+1}\times H_{\ell}}$
are the $\ell$th layer linear transformation matrices, 
$H_{\ell}$ is the hidden state size at the $\ell$th layer,
and
$g(\cdot)$ is the activation function,
such as a rectified linear unit, ${\rm ReLU}(\cdot)=\max(0,\cdot)$.
The parameters to be estimated are 
$\vec{W}=\{\vec{W}_{\rm in}^{(\ell)}$,
$\vec{W}_{\rm out}^{(\ell)}$,
$\vec{W}_{\rm self}^{(\ell)}\}_{\ell=0}^{L-1}$.
The last ($L$th) layer hidden state corresponds to the outputs,
$\vec{z}^{(L)}=f(\vec{G},\vec{x};\vec{W})$,
where $f$ represents the GCN.
Figure~\ref{fig:gcn} shows the calculation of the hidden state of the next layer.
By using different transformations for incoming and outcoming nodes with $\vec{W}_{\rm in}^{(\ell)}$ and
$\vec{W}_{\rm out}^{(\ell)}$, we can handle directed flow information on the graph,
which is important for traffic modeling~\cite{DBLP:journals/corr/LiYSL17}.
A graph convolutional layer (\ref{eq:gcn}) can collect information from nodes that are directly connected.
By using $L$ layers, information on nodes within the $L$-hop distance can be used for the output,
which is transformed so as to be beneficial for selecting a road to conduct the regulation.

For the loss function, we use
the Kullback-Leibler divergence
between the softmaxed regulation effect and
the outputs as follows,
\begin{align}
E &= \sum_{n=1}^{N}{\rm KL}({\rm softmax}(\vec{y}_{n})\parallel
{\rm softmax}(\vec{z}_{n})).
% \nonumber\\
%&\propto -\sum_{n=1}^{N}\sum_{i=1}^{I_{n}}
%\frac{\exp(y_{ni})}{\sum_{j}\exp(y_{nj})}
%\log \frac{\exp(z_{ni}^{(L)})}{\sum_{j}\exp(z_{nj}^{(L)})}.
\end{align}
By minimizing this loss function,
the regulations that have relatively high effects are likely to be high output values,
even if the regulation is not the best,
which leads to robust learning of the function.

\subsection{Regulation output}

Given road graph $\vec{G}$ and traffic situation $\vec{x}$,
a regulation is selected by using the learned function
$\vec{z}^{(L)}=f(\vec{G},\vec{x};\hat{\vec{W}})$ as follows,
$\hat{i}=\argmax_{i\in\{1,\cdots,I\}} z_{i}^{(L)}$,
which indicates road $\hat{i}$ is the road to close.
Here, $\hat{\vec{W}}$ is the estimated parameters of the GCN.
In some applications, we would like to rank the regulations according to their effects.
The ranking is obtained by $\vec{z}^{(L)}$.
If we have a simulator for the target situation,
by running simulations under a regulation in the ranked order,
we would find the best regulation with the smaller number of simulation runs.

\subsection{Extensions}
\label{sec:extension}

We considered closing a road for regulation.
However, the proposed method is applicable to other types of regulations
if they can be represented by a graph.
We can conduct multiple regulation operations,
e.g. closing several roads,
by applying the proposed method recursively;
1) finding a road to close, 2) modifying the road graph by closing the road,
and 3) finding another road to close.
If a simulator is available, the traffic situation can also be updated
by using the simulator when applying the selected regulation.

We considered the population on each road in terms of traffic situation.
Other features on each road,
such as
populations over time, average moving speed and road width,
can be used by including them in the input vector $\vec{x}=\vec{z}^{(0)}$ of the GCN.

\section{Experiments}
\label{sec:experiments}

\subsection{Data}

We evaluated the proposed method using real-world road graphs in Japan 
for finding a road to close to reduce the time taken to arrive at the destination.
We collected junction graphs around train stations
with an area of 1,450 meters square from the website of 
the Geographical Information Authority of Japan~\footnote{\url{https://maps.gsi.go.jp/development/ichiran.html}}.
Then, we constructed the road graphs from the junction graphs,
where the average number of nodes (roads) was 1,227, 
and the average number of edges (road connections) was 2,075.

We ran a 30-minute simulation with 100,000 pedestrians to obtain the traffic situation for each graph.
With each pedestrian, the train station was set as either the origin or the destination,
and these destinations or origins were randomly selected from junctions.
Then, we closed a road and ran 
an additional 30-minute simulation to calculate the 
regulation effect.
For the regulation effect, we used the average reduction in the time taken to arrive at the destination.
When there was no route to the destination as the result of closing a road,
we set the regulation effect as the worst value in that situation.
With 19\% of the roads the arrival time was sooner with closure than
that without the regulation.
With 48\% of the roads the arrival time was later, and there was
no effect for 33\% of the roads.
The average arrival time was 704.9 seconds, and
the average maximum time reduction was 76.3 seconds.
The number of road graphs for training, validation and test were 1,500, 100, and 100, respectively.
Note that road graphs in the test data were not included in the
training and validation data, 
and we needed to find a regulation for unseen areas.

\subsection{Simulator}

To obtain traffic situations and regulation effects, 
we used an inhouse crowd simulator based on a multi-agent system. % as shown in Figure~\ref{fig:simulator}.
With the simulator,
each agent moves to the destination along the shortest path 
given the origin, destination, start time and road graph.
The maximum speed $v_{\rm max}$ is drawn from
a log normal distribution with a median of 1.16 meter per second
and a standard deviation of 0.12 for each agent.
Each agent moves at the maximum speed if there is no congestion,
and the speed decreases with the degree of congestion.
Specifically, the speed $v$ is determined by
\begin{align}
v=
\begin{cases}
v_{\rm max} & 0\leq \rho < \frac{v_{\rm max}+0.3}{1.8}\\
1.8\rho^{-1}-0.3 & \frac{v_{\rm max}+0.3}{1.8} \leq \rho \leq 6\\
0 & \rho \geq 6,
\end{cases}
\end{align}
%$v=v_{\rm max}$ $(0\leq \rho < \frac{v_{\rm max}+0.3}{1.8})$,
%$v=1.8\rho^{-1}-0.3$ $(\frac{v_{\rm max}+0.3}{1.8} \leq \rho \leq 6)$,
%and $v=0$  $(\rho \geq 6)$,
where $\rho$ is the population density in front of the agent.

%\begin{figure}
%\centering
%\includegraphics[width=21em]{outputSS.png} 
%\caption{Simulator used in our experiments.}
%\label{fig:simulator}
%\end{figure}

\subsection{Evaluation measurements}

We evaluated the proposed method using the following three evaluation measurements:
time reduction, top-10 accuracy, and maximum time reduction through $K$ simulations.
The time reduction is the average reduction in the time 
taken to arrive at the destination 
compared with the arrival time without regulation.
The time reduction is positive when the arrival time is reduced by the regulation,
and it is zero when the arrival time is the same with that without regulation.
The top-10 accuracy is a rate when the best regulation is included in the top-10 list.
The best regulation is that with the maximum time reduction of all regulations. 
With the maximum time reduction through $K$ simulations, we assume that we can use simulations
to check suggested regulations.
By conducting simulations with different regulations in a ranked order,
we can find the best regulation according to the maximum time reduction.
Since we would like to find a better regulation in a shorter time,
a larger time reduction with fewer simulations is preferred.

\subsection{Comparing methods}

We compared the proposed method with the Population, Closeness, Betweenness, Bayesian optimization, and Random methods.
When ranking the roads to close, the Population method uses the road information,
the Closeness and Betweenness methods use the road graph structures,
and the Bayesian optimization method uses simulation results.
The Random method randomly ranks the roads.

The Population method ranks roads to close in descending order of population of each road.
It is intended to disperse people in a congested road.

The Closeness method ranks roads to close in descending order of closeness centrality~\cite{bavelas1950communication}.
A road that can access other roads with short path lengths has high closeness centrality.
The closeness centrality is calculated as the inverse of the sum of the shortest path length between the node (road)
and all the other nodes in the graph as follows,
${\rm closeness}_{i}=\frac{1}{\sum_{j\neq i}d(i,j)}$,
where $d(i,j)$ is the shortest path length between nodes $i$ and $j$.

The Betweenness method ranks roads to close in descending order of betweenness centrality~\cite{freeman1977set}.
A road that is often used by traffic between all pairs of nodes has high betweenness centrality.
The betweenness centrality is calculated as the sum of the fraction of all-pair shortest paths 
that pass through the node as follows,
${\rm betweenness}_{i}=\sum_{j,k}\frac{\sigma_{jk}(i)}{\sigma_{jk}}$,
where $\sigma_{jk}$ is the number of shortest paths between nodes $j$ and $k$,
and $\sigma_{jk}(i)$ is the number of those paths passing through node $i$ other than $j$ and $k$.

The Bayesian optimization (BO) method
optimizes a black-box function with as few function evaluations 
as possible~\cite{pelikan1999boa}.
%as possible~\cite{pelikan1999boa,snoek2012practical,brochu2010tutorial}.
Since BO is applicable when function evaluations, 
in this case simulation runs, are possible,
we included BO as a comparative method for the maximum time reduction through $K$ simulations,
but not for the time reduction and top-10 accuracy.
With BO, a Gaussian process is used to approximate the time reduction
with the following two kernels: the RBF kernel on the population and the diffusion kernel on the road graph.
The RBF kernel is defined by
$k_{\rm RBF}(i,j)=\alpha \exp\left(-\frac{\gamma}{2}\parallel x_{i}-x_{j}\parallel^{2}\right)+\beta\delta_{i,j}$,
where $\alpha$, $\beta$ and $\gamma$ are the hyperparameters,
$\delta_{ij}$ is the Kronecker delta, $\delta_{ij}=1$ if $i=j$ and $\delta_{ij}=0$ otherwise,
and $x_{i}$ is the population on road $i$.
The diffusion kernel is defined by
$\vec{K}_{\rm diffusion}=\exp(-\eta\vec{L})$,
where $\eta$ is the hyperparameter, and $\vec{L}$ is the graph Laplacian of the road graph.
We selected the hyperparameters by employing a grid search.

The proposed method used
the GCN with $L=10$ layers of $H_{\ell}=10$ hidden states on each layer.
The activation function was ReLU, and it was followed by
batch normalization. 
%~\cite{ioffe2015batch}.
The best model was selected by using validation data 
based on the time reduction.
The GCN was implemented by using a deep learning framework, Chainer~\cite{tokui2015chainer}.

\subsection{Results}

The time reduction and top-10 accuracy shown in Table~\ref{tab:result}
were averaged over ten experiments 
with different training, validation and test data sets.
The proposed method successfully reduced the average arrival time,
and achieved the maximum time reduction.
On the other hand, the time reduction with the other methods were negative;
they increased the average arrival time.
The top-10 accuracy with the proposed method was higher 
than that with the other methods.
These results indicates that the proposed method
can determine a regulation operation by learning a function using GCNs.

\begin{table}[t!]
\centering
\caption{Average time reduction and top-10 accuracy.}
\label{tab:result}
\begin{tabular}{lrr}
\hline
& Time reduction & Top-10 accuracy \\
\hline
Proposed & ${\bf 9.93 \pm 2.12}$ & {\bf 0.256 $\pm$ 0.014} \\
Random & $-29.25 \pm 1.85$ & 0.013 $\pm$ 0.004\\
Population & $-37.57 \pm 1.81$ & 0.105 $\pm$ 0.010\\
Closeness & $-5.89 \pm 1.08$ & 0.111 $\pm$ 0.010\\
Betweenness & $-23.07 \pm 1.94$ & 0.107 $\pm$ 0.010\\
\hline
\end{tabular} 
\end{table}

\begin{figure}[t!]
\centering
\includegraphics[width=23em]{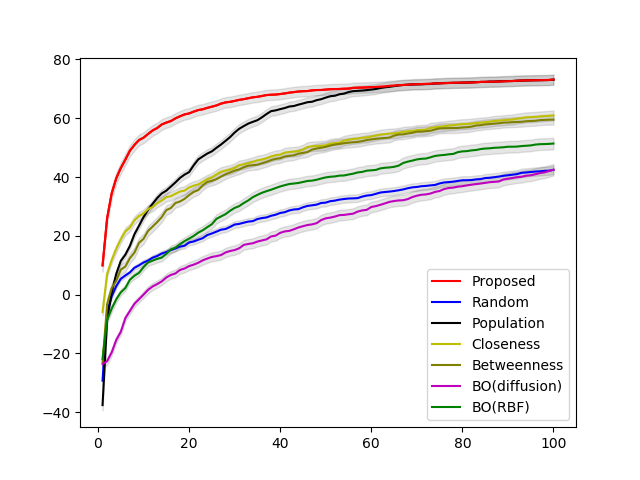} 
\caption{Maximum time reduction through $K$ simulations. The horizontal axis is the number of simulations $K$, and the vertical axis is the maximum time reduction.}
\label{fig:bo}
\end{figure}

Figure~\ref{fig:bo} shows the maximum time reduction through
$K$ simulations.
The proposed method achieved the highest maximum time reduction
over every number of simulation runs.
With the proposed method, the maximum time reduction quickly increased
in a small number of simulation runs.
This result indicates that we can obtain better regulations
by running simulations in the ranked order by the proposed method.
The Population method gave the lowest maximum time reduction without simulations,
but the time reduction increased by running simulations.
The maximum time reduction with the Bayesian optimization methods, BO(diffusion) and BO(RBF), was small. This is
because the time reduction over the road graph was
difficult to approximate by Gaussian processes.

\begin{figure*}[t!]
\centering
 \begin{tabular}{ccc}
& Area1 &\\
%\tfbox{\includegraphics[width=16em]{station3972_method0.png}}&
\tfbox{\includegraphics[width=16em]{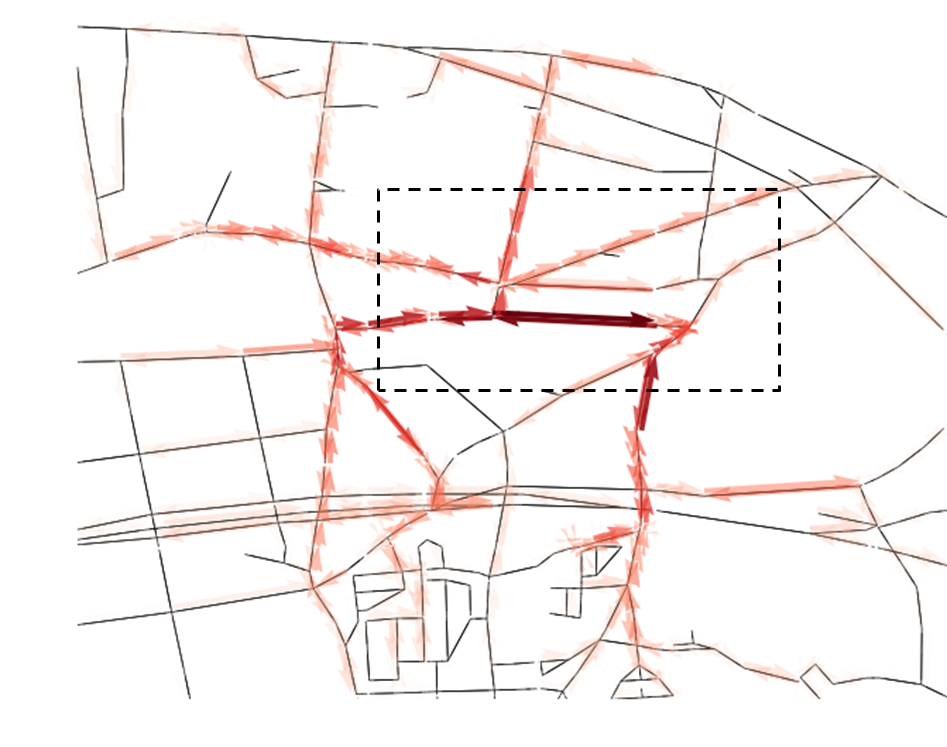}}&
\tfbox{\includegraphics[width=16em]{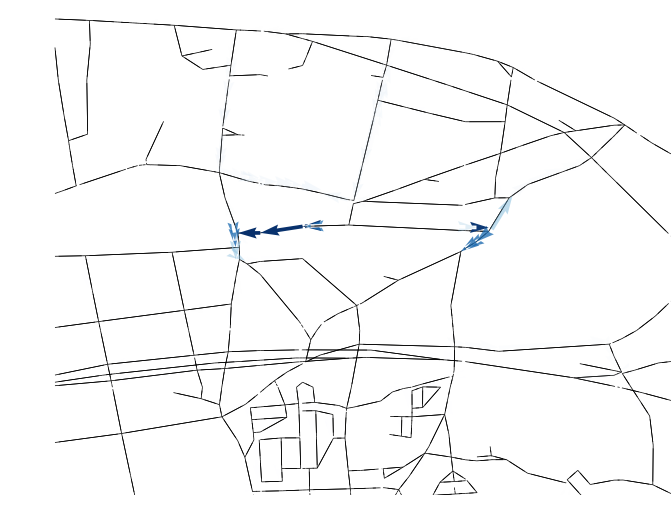}}&
\tfbox{\includegraphics[width=16em]{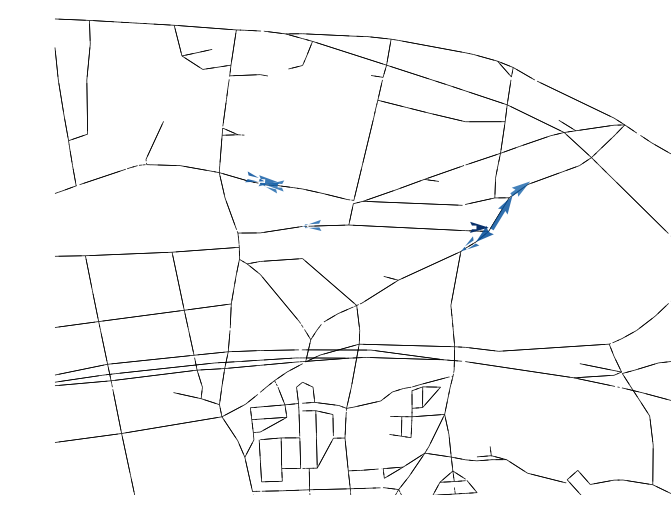}}\\
& Area2 &\\
\tfbox{\includegraphics[width=16em]{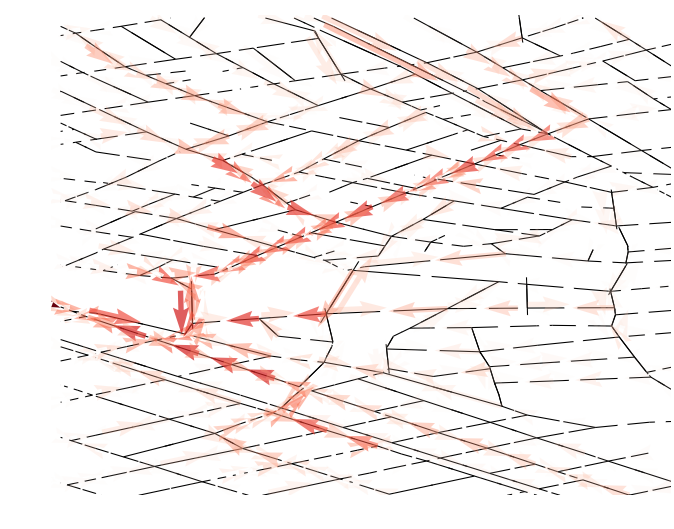}}
&
\tfbox{\includegraphics[width=16em]{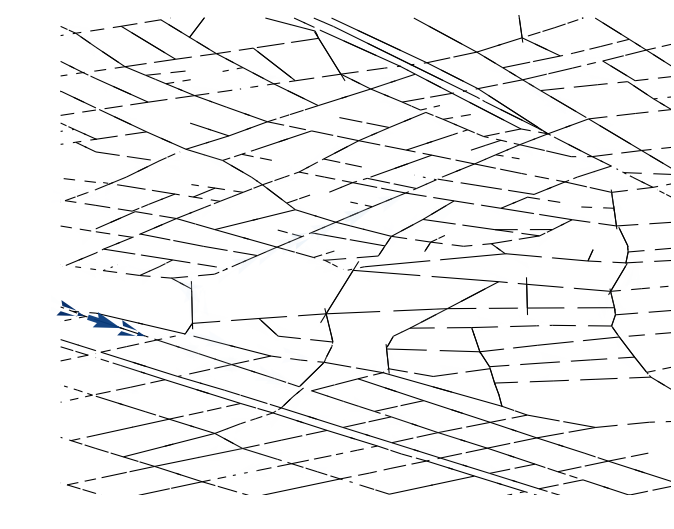}}&
\tfbox{\includegraphics[width=16em]{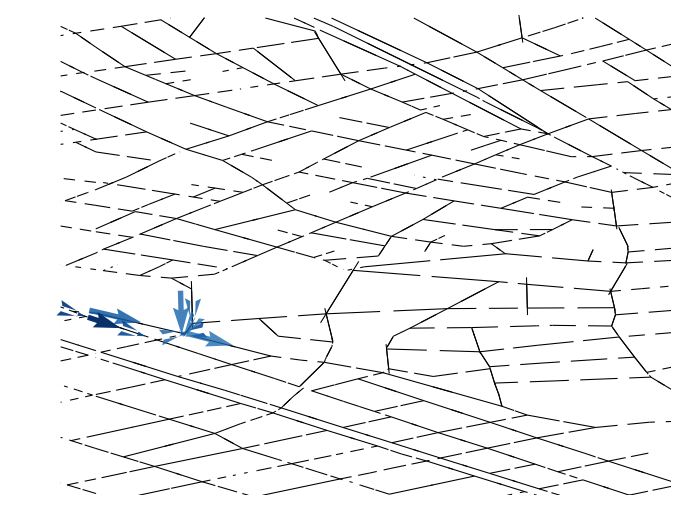}}\\
(a) Population of each road
&
(b) True regulation effects
&
(c) Estimated regulation effects
\end{tabular}
\caption{(a) Population of each road, (b) roads with truly high regulation effects which are calculated by using simulations, and (c) roads with high estimated regulation effects estimated by the proposed method. A darker arrow indicates a higher value. A black line represents a road. }
\label{fig:map}
\end{figure*}

\begin{figure*}[t!]
\centering
{\tabcolsep=2em
\begin{tabular}{cc}
\tfbox{\includegraphics[height=9em]{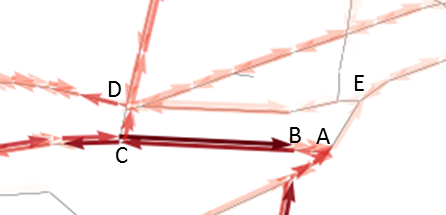}}&
\tfbox{\includegraphics[height=9em]{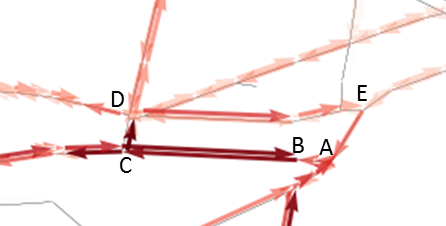}}\\
(a) Without regulation &
(b) With regulation by the proposed method\\
\end{tabular}}
\caption{(a) Population without regulation, and (b) population when road $B \to A$ is closed.
The map shows the boxed area in Area1 of Figure~\ref{fig:map}(a).}
\label{fig:navi}
\end{figure*}

Figure~\ref{fig:map} shows the population of each road, true regulation effects 
and estimated regulation effects obtained with 
the proposed method for two example areas.
Truly high regulation effect roads were successfully included 
in the roads with high estimated regulation effects 
estimated by the proposed method.

Figure~\ref{fig:navi} shows the population for a one-hour simulation time
without regulation (a), and the population for a one-hour simulation time with regulation with a 30-minute simulation time (b), where 
the proposed method selected road $B \to A$ to close as the regulation.
The population on road $C \to B$ was large.
Without regulation, people who go to $E$ from $C$ also use the road, and the road was heavily congested.
By closing road $B \to A$,
people who went to $E$ used a different route $C \to D \to E$, 
Then, the congestion on road $C \to B$ was alleviated,
and the average arrival time was reduced.

\section{Related work}
\label{sec:related}

Simulations have been used to aid crowd regulations~\cite{DBLP:conf/huc/UedaNSIOS15}.
However, this approach requires data assimilation and multiple simulation runs.
Although Bayesian optimization~\cite{pelikan1999boa},
%~\cite{pelikan1999boa,snoek2012practical,brochu2010tutorial},
which is a method for optimizing of black-box functions,
can reduce the number of simulation runs,
data assimilation and multiple simulation runs are still needed.
A number of methods have been proposed for adaptive crowd regulations, 
especially signal control, using reinforcement 
learning~\cite{choi1996predictive,wiering2000multi,khamis2014adaptive,genders2016using,gao2017adaptive}.
However, these methods consider a specific area, and learned models cannot be used for other areas.
The proposed method can find regulations on any areas by using road graph information
without data assimilation or simulation runs.
Regulation methods for individuals, i.e. cars and pedestrians,
have been extensively studied~\cite{yamashita2005smooth,lin2017real,may2003pedestrian,arikawa2007navitime}.
These methods find a route with the minimum travel time for each individual 
given the origin and destination.
The proposed method finds a regulation for a crowd, and does not use origin and destination information.
Graph convolutional networks have been successfully used 
for a wide variety of applications, 
such as 
chemicals~\cite{duvenaud2015convolutional},
semi-supervised learning~\cite{kipf2016semi}
and traffic forecasting~\cite{DBLP:journals/corr/abs-1709-04875,DBLP:journals/corr/LiYSL17},
but not for finding traffic regulations.

%~\cite{defferrard2016convolutional,duvenaud2015convolutional,pmlr-v70-gilmer17a,li2015gated,schutt2017moleculenet,schlichtkrull2017modeling,seo2016structured,kipf2016semi,DBLP:journals/corr/abs-1709-04875,DBLP:journals/corr/LiYSL17}, but not for regulation.
%Since the proposed method uses a deep learning method to solve an optimization problem,
%it is related to methods for learning optimization algorithms using 
%recurrent or feedforward neural networks~\cite{wang2016learning,xin2016maximal,andrychowicz2016learning,gregor2010learning,sprechmann2015learning,he2017from}.
%The proposed method utilizes data generated by simulators for supervised learning.
%Such approach has been used for human detection~\cite{buys2013virtual}
%and gaze estimation~\cite{wood2015rendering}, where realistic simulators are available.
%chemicals~\cite{duvenaud2015convolutional}
%traffic networks~\cite{DBLP:journals/corr/abs-1709-04875,DBLP:journals/corr/LiYSL17},
%knowledge graphs~\cite{schlichtkrull2017modeling}
%semi-supervised learning~\cite{kipf2016semi}
%and text documents~\cite{seo2016structured}.

\section{Conclusion}
\label{sec:conclusion}

In this paper, we proposed a method for finding appropriate regulations.
With the proposed method, a function that outputs regulation effects given traffic situations and road graphs
is modeled with graph convolutional networks.
With experiments using real-world road graphs and a crowd simulator,
we confirmed that the proposed method can find appropriate regulations efficiently.
Although our results have been encouraging, our framework can be further improved upon in a number of ways. 
Firstly, we plan to use time information by incorporating recurrent neural networks.
Secondly, we would like to extend our framework to adaptive regulation by using reinforcement learning. 
%Thirdly, the proposed method can incorporate learning to rank techniques~\cite{burges2005learning},
%which enables us to learn functions without the need of preparing all regulation patters.

\bibliographystyle{abbrv}

%\begin{small}
%\bibliography{icdm2018}

\begin{thebibliography}{10}

\bibitem{almeida2013crowd}
J.~E. Almeida, R.~J. Rosseti, and A.~L. Coelho.
\newblock Crowd simulation modeling applied to emergency and evacuation
  simulations using multi-agent systems.
\newblock {\em arXiv preprint arXiv:1303.4692}, 2013.

\bibitem{arikawa2007navitime}
M.~Arikawa, S.~Konomi, and K.~Ohnishi.
\newblock Navitime: Supporting pedestrian navigation in the real world.
\newblock {\em IEEE Pervasive Computing}, 6(3), 2007.

\bibitem{bavelas1950communication}
A.~Bavelas.
\newblock Communication patterns in task-oriented groups.
\newblock {\em The Journal of the Acoustical Society of America},
  22(6):725--730, 1950.

\bibitem{braun2003modeling}
A.~Braun, S.~R. Musse, L.~P.~L. de~Oliveira, and B.~E. Bodmann.
\newblock Modeling individual behaviors in crowd simulation.
\newblock In {\em Computer Animation and Social Agents, 2003. 16th
  International Conference on}, pages 143--148. IEEE, 2003.

\bibitem{choi1996predictive}
S.~P. Choi and D.-Y. Yeung.
\newblock Predictive {Q}-routing: A memory-based reinforcement learning
  approach to adaptive traffic control.
\newblock In {\em Advances in Neural Information Processing Systems}, pages
  945--951, 1996.

\bibitem{duvenaud2015convolutional}
D.~K. Duvenaud, D.~Maclaurin, J.~Iparraguirre, R.~Bombarell, T.~Hirzel,
  A.~Aspuru-Guzik, and R.~P. Adams.
\newblock Convolutional networks on graphs for learning molecular fingerprints.
\newblock In {\em Advances in Neural Information Processing Systems}, pages
  2224--2232, 2015.

\bibitem{freeman1977set}
L.~C. Freeman.
\newblock A set of measures of centrality based on betweenness.
\newblock {\em Sociometry}, pages 35--41, 1977.

\bibitem{gao2017adaptive}
J.~Gao, Y.~Shen, J.~Liu, M.~Ito, and N.~Shiratori.
\newblock Adaptive traffic signal control: Deep reinforcement learning
  algorithm with experience replay and target network.
\newblock {\em arXiv preprint arXiv:1705.02755}, 2017.

\bibitem{genders2016using}
W.~Genders and S.~Razavi.
\newblock Using a deep reinforcement learning agent for traffic signal control.
\newblock {\em arXiv preprint arXiv:1611.01142}, 2016.

\bibitem{khamis2014adaptive}
M.~A. Khamis and W.~Gomaa.
\newblock Adaptive multi-objective reinforcement learning with hybrid
  exploration for traffic signal control based on cooperative multi-agent
  framework.
\newblock {\em Engineering Applications of Artificial Intelligence},
  29:134--151, 2014.

\bibitem{kipf2016semi}
T.~N. Kipf and M.~Welling.
\newblock Semi-supervised classification with graph convolutional networks.
\newblock {\em arXiv preprint arXiv:1609.02907}, 2016.

\bibitem{DBLP:journals/corr/LiYSL17}
Y.~Li, R.~Yu, C.~Shahabi, and Y.~Liu.
\newblock Diffusion graph convolutional recurrent neural network: Data-driven
  traffic forecasting.
\newblock {\em CoRR}, abs/1707.01926, 2017.

\bibitem{lin2017real}
J.~Lin, W.~Yu, X.~Yang, Q.~Yang, X.~Fu, and W.~Zhao.
\newblock A real-time en-route route guidance decision scheme for
  transportation-based cyberphysical systems.
\newblock {\em IEEE Transactions on Vehicular Technology}, 66(3):2551--2566,
  2017.

\bibitem{may2003pedestrian}
A.~J. May, T.~Ross, S.~H. Bayer, and M.~J. Tarkiainen.
\newblock Pedestrian navigation aids: information requirements and design
  implications.
\newblock {\em Personal and Ubiquitous Computing}, 7(6):331--338, 2003.

\bibitem{murakami2002multi}
Y.~Murakami, K.~Minami, T.~Kawasoe, and T.~Ishida.
\newblock Multi-agent simulation for crisis management.
\newblock In {\em Knowledge Media Networking, 2002. Proceedings. IEEE Workshop
  on}, pages 135--139. IEEE, 2002.

\bibitem{pelikan1999boa}
M.~Pelikan, D.~E. Goldberg, and E.~Cant{\'u}-Paz.
\newblock {BOA}: The {B}ayesian optimization algorithm.
\newblock In {\em Proceedings of the 1st Annual Conference on Genetic and
  Evolutionary Computation-Volume 1}, pages 525--532. Morgan Kaufmann
  Publishers Inc., 1999.

\bibitem{schlichtkrull2017modeling}
M.~Schlichtkrull, T.~N. Kipf, P.~Bloem, R.~van~den Berg, I.~Titov, and
  M.~Welling.
\newblock Modeling relational data with graph convolutional networks.
\newblock {\em arXiv preprint arXiv:1703.06103}, 2017.

\bibitem{tokui2015chainer}
S.~Tokui, K.~Oono, S.~Hido, and J.~Clayton.
\newblock Chainer: a next-generation open source framework for deep learning.
\newblock In {\em Proceedings of Workshop on Machine Learning Systems in the
  Annual Conference on Neural Information Processing Systems}, volume~5, 2015.

\bibitem{DBLP:conf/huc/UedaNSIOS15}
N.~Ueda, F.~Naya, H.~Shimizu, T.~Iwata, M.~Okawa, and H.~Sawada.
\newblock Real-time and proactive navigation via spatio-temporal prediction.
\newblock In {\em Proceedings of the 2015 {ACM} International Joint Conference
  on Pervasive and Ubiquitous Computing and Proceedings of the 2015 {ACM}
  International Symposium on Wearable Computers}, pages 1559--1566, 2015.

\bibitem{wiering2000multi}
M.~Wiering.
\newblock Multi-agent reinforcement learning for traffic light control.
\newblock In {\em Machine Learning: Proceedings of the Seventeenth
  International Conference (ICML'2000)}, pages 1151--1158, 2000.

\bibitem{yamashita2005smooth}
T.~Yamashita, K.~Izumi, K.~Kurumatani, and H.~Nakashima.
\newblock Smooth traffic flow with a cooperative car navigation system.
\newblock In {\em Proceedings of the fourth international joint conference on
  Autonomous agents and multiagent systems}, pages 478--485. ACM, 2005.

\bibitem{youn2008price}
H.~Youn, M.~T. Gastner, and H.~Jeong.
\newblock Price of anarchy in transportation networks: efficiency and
  optimality control.
\newblock {\em Physical review letters}, 101(12):128701, 2008.

\bibitem{DBLP:journals/corr/abs-1709-04875}
B.~Yu, H.~Yin, and Z.~Zhu.
\newblock Spatio-temporal graph convolutional neural network: {A} deep learning
  framework for traffic forecasting.
\newblock {\em CoRR}, abs/1709.04875, 2017.

\end{thebibliography}
%\end{small}

\end{document}